\documentclass[twocolumn]{aa}
\usepackage{graphicx}
\usepackage{txfonts}
\begin{document}
\def\lsim{\,\lower2truept\hbox{${< \atop\hbox{\raise4truept\hbox{$\sim$}}}$}\,}
\def\gsim{\,\lower2truept\hbox{${> \atop\hbox{\raise4truept\hbox{$\sim$}}}$}\,}

   \title{An angular power spectrum analysis of the
          DRAO 1.4 GHz polarization survey: implications for CMB
       observations}

   \author{L.~La~Porta\inst{1,}\thanks{Member of the
          International Max Planck Research Shool (IMPRS)
for Radio and Infrared Astronomy at the Universities
    of Bonn and Cologne.},
          C.~Burigana\inst{2},
   W.~Reich\inst{1},
   \and
   P.~Reich\inst{1}
          }

   \offprints{C.~Burigana}

   \institute{Max-Planck-Institut f\"ur Radioastronomie,
              Auf dem H\"ugel, 69, D-53121 Bonn, Germany\\
             \and
              INAF-IASF Bologna,
              via P.~Gobetti, 101, I-40129 Bologna, Italy\\
             }

   \date{Submitted May 3, 2006; in revised form May 26, 2006.}

   \abstract{}
{The aim of the present analysis is to improve the knowledge of the statistical properties
of the Galactic diffuse synchrotron emission, which
constrains sensitive
CMB anisotropy measurements.}
{We have analysed the new DRAO 1.4 GHz polarization survey
together with the Stockert 1.4 GHz total intensity survey
 and derived the angular power spectra (APSs) of the
 total intensity, the polarized emission,
and their cross-correlation for the entire surveys and for three
low-intensity regions. }
   {The APSs of the diffuse synchrotron emission
   are modelled by power laws. For the $E$
   and $B$ modes, a slope of $\alpha \sim [-3.0,-2.5]$
   for the multipole range $\sim [30,300]$ is found.
   By the extrapolation of these results to 70~GHz, we can estimate
   the Galactic synchrotron contamination of CMB anisotropies,
and we find results that are compatible with the ones coming
from WMAP 3-yr data.
In the low-intensity regions,
 the  cosmological primordial B~mode peak at $\ell \sim 100$  should be
   clearly observable for a tensor-to-scalar ratio $T/S \gsim 0.5$
   and a synchrotron temperature spectral index $\beta \sim -3$.
    Its detection is also possible for
   $T/S \gsim 0.005$ and $\beta \sim -3$,
  in case a separation of the foreground from the CMB signal
   could be achieved with an accuracy of $\sim 5-10\%$.
     For the TE mode, a mask
   excluding $|b_{gal}| \le 5^{\circ}$ (for $\beta\sim -3.0$)
   or $|b_{gal}| \le 20^{\circ}$ (for $\beta\sim -2.8$)
   from the surveys is sufficient to render
   the foreground contamination negligible, thus confirming the ability
   of WMAP to have a clear view of the temperature-polarization
   correlation peak and antipeak series.
  }{}


\authorrunning{L.~La~Porta et al.}
\titlerunning{APS of the Galactic synchrotron emission at 1.4 GHz}

   \maketitle
%

\section{Introduction}

The polarized Galactic diffuse synchrotron radiation is expected
to be the major foreground at $\nu \lsim 70$~GHz on intermediate
and large angular scales ($\theta \gsim 30'$)
at medium and high Galactic latitudes.
At about 1~GHz, the synchrotron emission is the most important radiative
mechanism out of the Galactic plane.
Consequently, radio frequencies are the natural range for studying it,
though it might be affected by Faraday rotation and depolarization.
Before the DRAO survey became available, the Leiden polarization
surveys (\cite{spo76})
provided a
sky coverage,
allowing to estimate the Galactic synchrotron APS on intermediate
and large scales (\cite{lavspoelstra}).
However, these surveys have sparse sampling,
low sensitivity, and a good signal-to-noise ratio only for
the brightest regions in the sky.
A new linear polarization survey
of the northern celestial
hemisphere at 1.42 GHz with
an angular resolution
$\theta_{HPBW} \simeq 36\arcmin$
has recently been completed
using the DRAO 26~m telescope
(\cite{wolleben06}).
The survey
has a spacing, $\theta_s$, of $15'$ in right ascension
and from $0\fdg25$ to $2\fdg5$
in declination,
which requires interpolation to construct equidistant cylindrical
({\tt ECP}) maps with $\theta_{pixel} = 15\arcmin$ (\cite{wollebenPhD}).
The final map has an rms-noise of about 12~mK, which is
unique so far
in terms of sensitivity.
The polarized intensity map appears cold and patchy
at high Galactic latitudes,
except for the ``North Polar Spur'' (NPS), extending far
out of the plane at about $\ell_{gal} \sim 30\degr$\footnote{The NPS
has been extensively studied (
see Egger \& Aschenbach~(1995) and references therein)
and is interpreted as the shock
front of an evolved local supernova remnant.}.
We used the polarization data in combination with the
Stockert total intensity map at 1.42 GHz (Reich~1982; Reich \& Reich~1986),
having the same angular resolution, similar sensitivity, and $\theta_s \simeq 15'$,
to investigate
the statistical properties of the Galactic diffuse synchrotron emission.
The major purpose of the analysis is the comparison of
the foreground fluctuation properties with those of the CMB
anisotropy field, as drawn in standard cosmologies. Therefore,
we described the anisotropy field statistical
properties in terms of its angular power spectrum (APS)\footnote{
The angular scale $\theta$ and the
angular power spectrum multipole $\ell$ are related by
$\ell \sim 180^{\circ}/\theta[^{\circ}]$.}
(see \cite{peebles};~\cite{kamionkowski97};~\cite{zaldarriaga01})
 and adopted  {\tt HEALPix}\footnote{http://healpix.jpl.nasa.gov/}
(\cite{gorski05})
as the tessellation scheme for the sphere.
A detailed description of the algorithm
to convert the original {\tt ECP} maps into
{\tt HEALPix} maps was given by La~Porta et al.~(2005).

  \begin{figure}[t]
\hskip -0.2cm
   \begin{tabular}{ccc}
   \includegraphics[bb = 118 85 508 537,width=2.6cm,height=3.cm,angle=0,clip=]{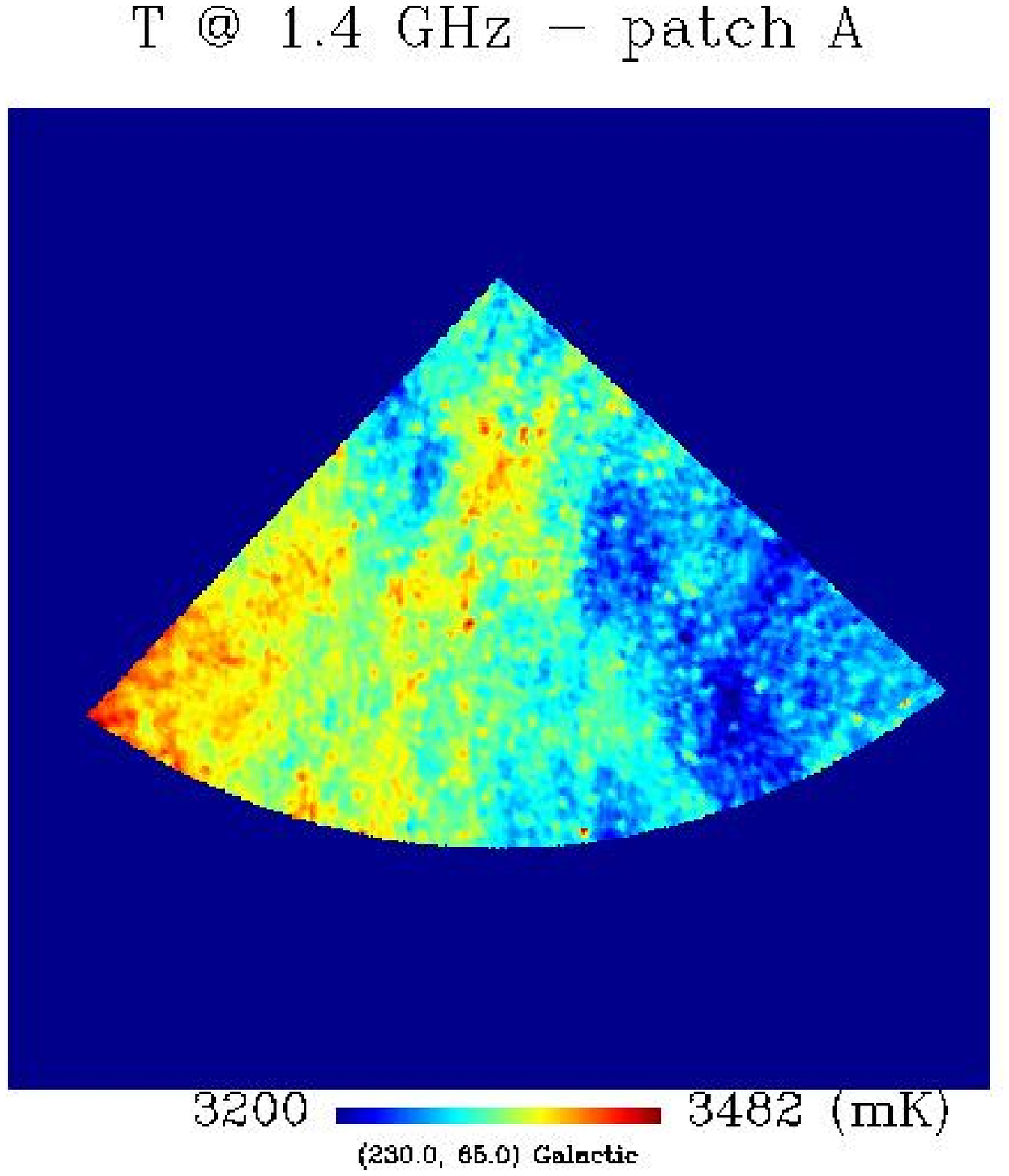}&
   \includegraphics[bb = 118 85 508 537,width=2.6cm,height=3.cm,angle=0,clip=]{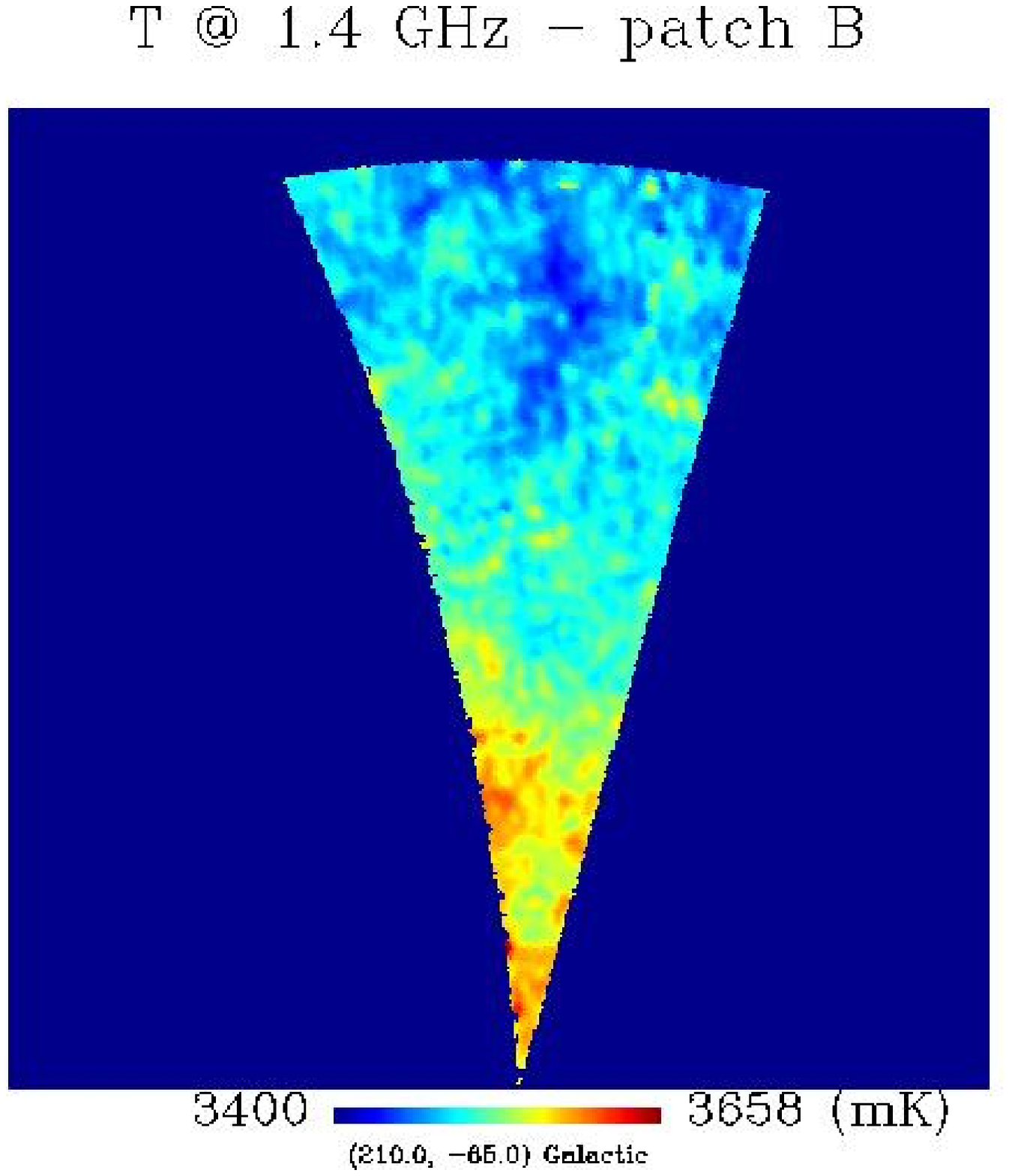}&
   \includegraphics[bb = 118 85 508 537,width=2.6cm,height=3.cm,angle=0,clip=]{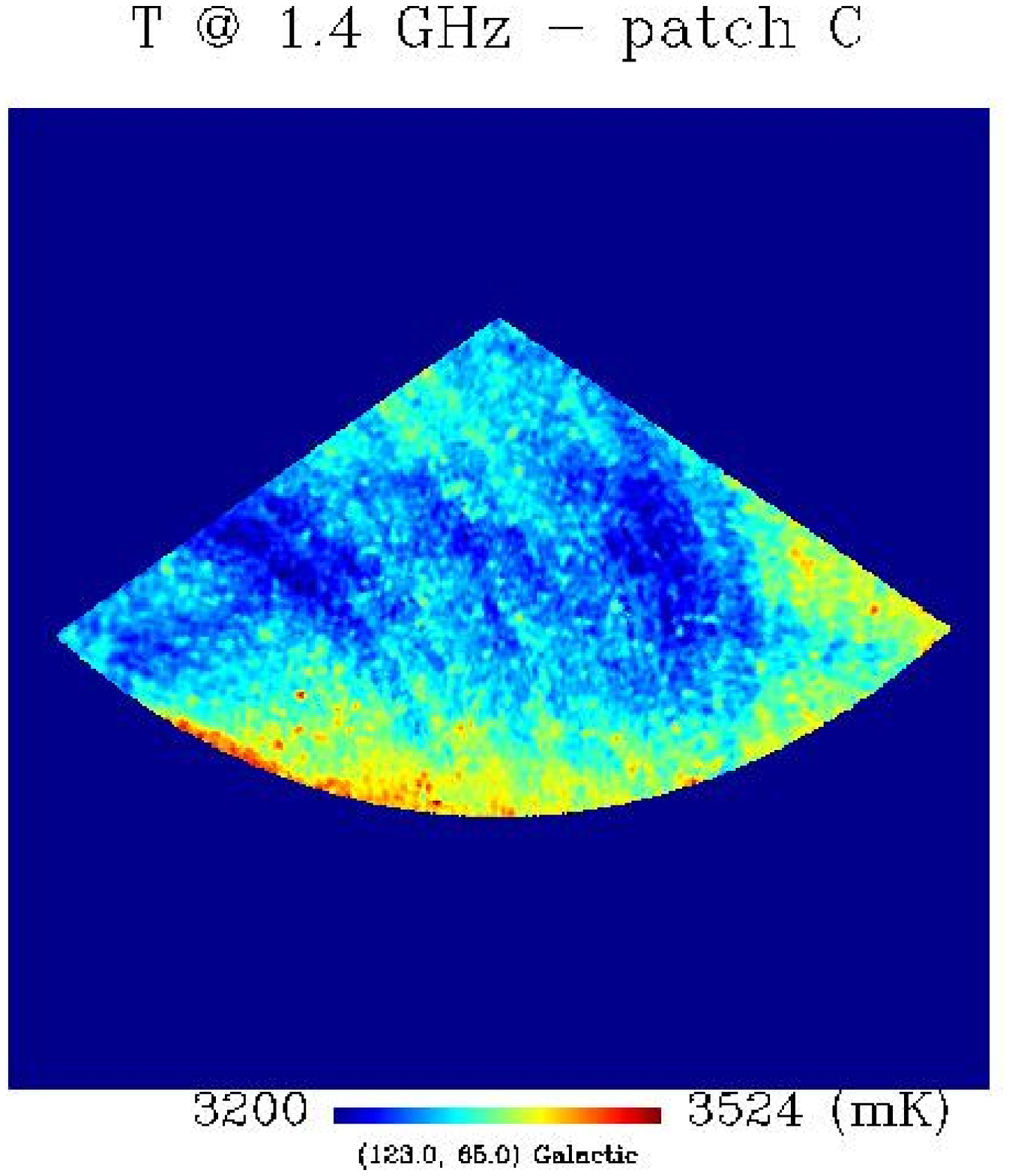}\\
   \includegraphics[bb = 118 85 508 537,width=2.6cm,height=3.cm,angle=0,clip=]{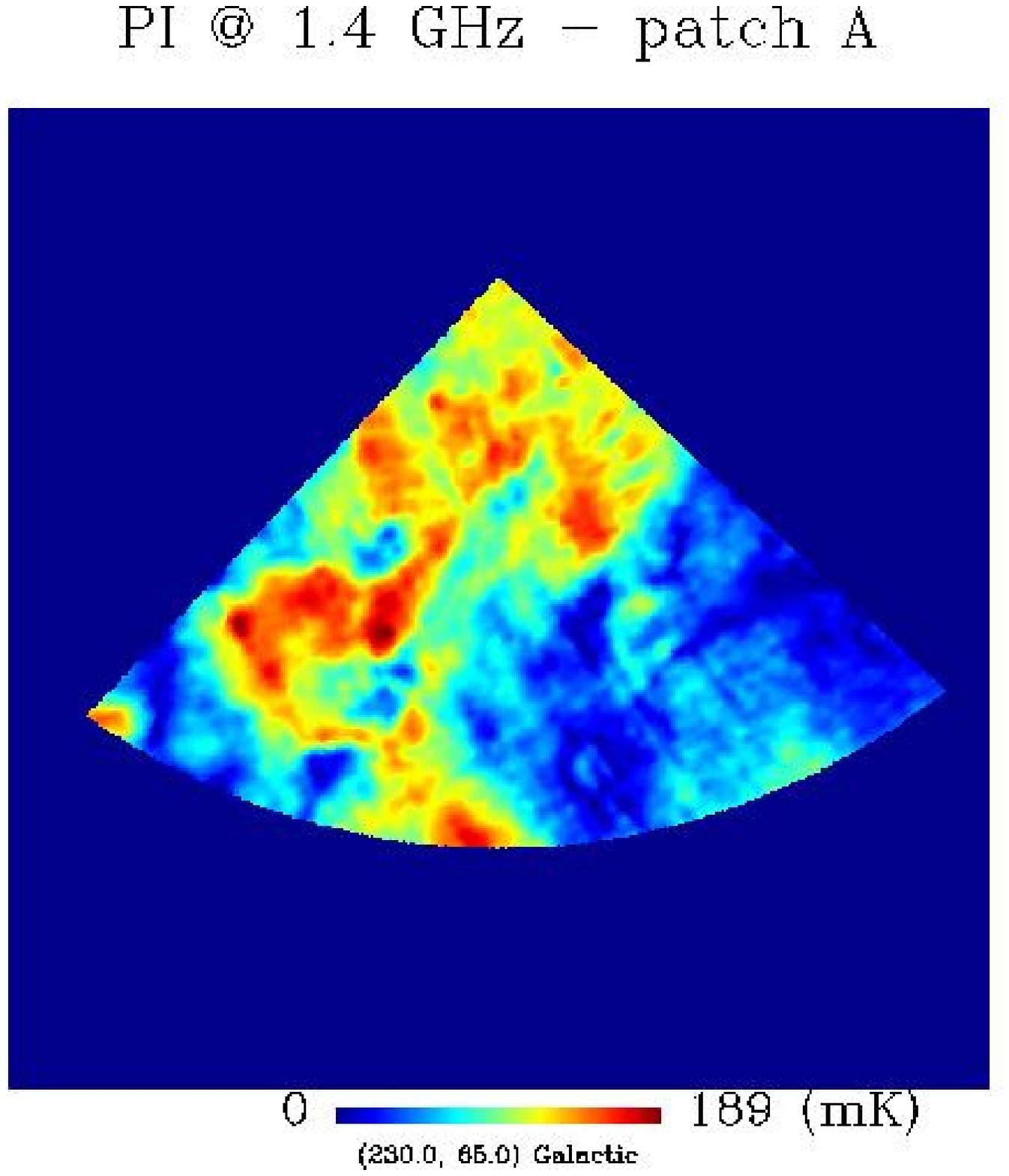}&
   \includegraphics[bb = 118 85 508 537,width=2.6cm,height=3.cm,angle=0,clip=]{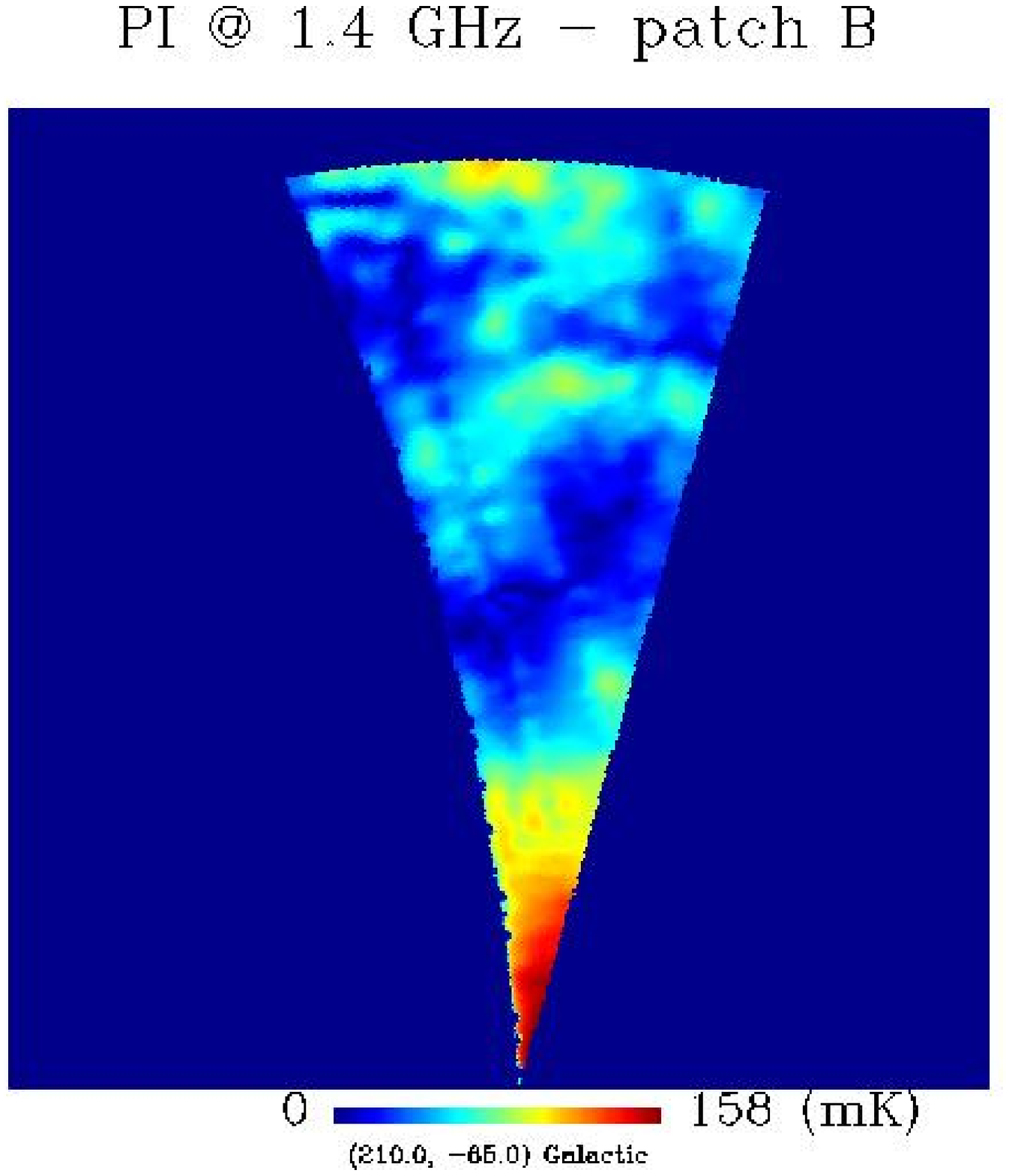}&
   \includegraphics[bb = 118 85 508 537,width=2.6cm,height=3.cm,angle=0,clip=]{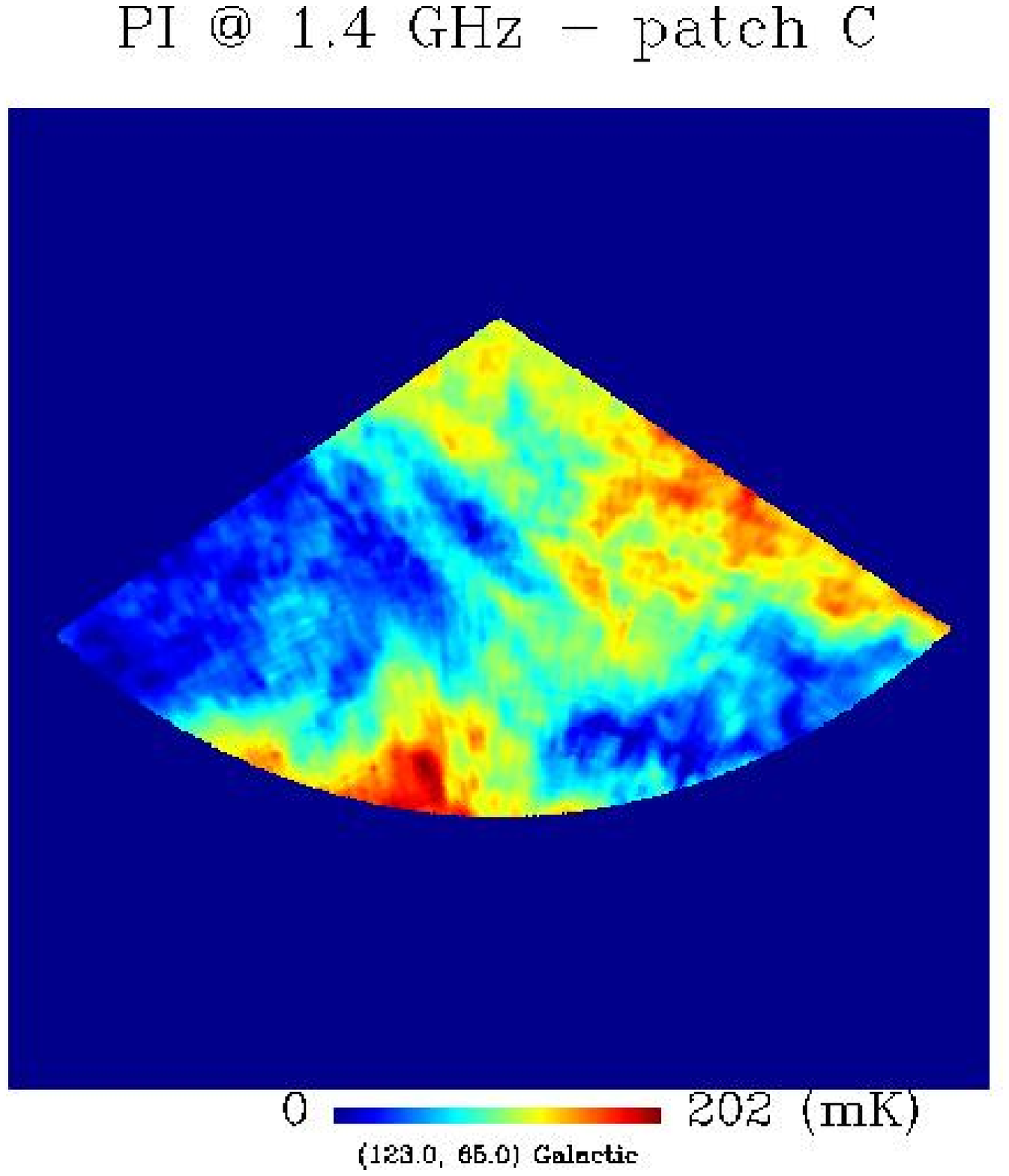}\\
   \end{tabular}
\vskip -0.4cm
   \caption{Gnomonic projection of the cold patches 
   representative of the diffuse Galactic synchrotron emission at 1.4 GHz
   (total intensity -- top panels; polarization -- bottom panels).
   }
   \label{patches}
 \end{figure}

\section{Selected areas and data analysis}

\begin{figure}[!th]
\hskip +0.7cm
  \includegraphics[width=3.7cm,height=8.3cm,angle=90]{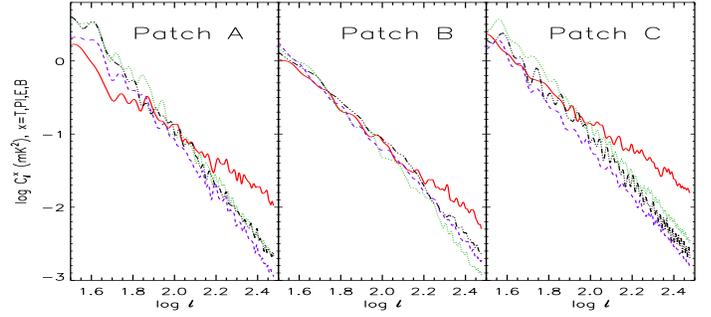}
\vskip +0.2cm
  \caption{APSs 
  of the selected cold regions 
  (solid lines $\to C_{\ell}^{TI}$; dashes $\to C_{\ell}^{PI}$;
  dots $\to C_{\ell}^{E}$; three dots-dashes $\to C_{\ell}^{B}$).
  }
  \label{lowTpatchaps1}
\end{figure}

We selected three areas with low polarized intensities
(see Fig.~\ref{patches}), which are
identified by the following Galactic coordinates:
Patch A - $180^{\circ} \le \ell_{gal} \le 276^{\circ}$, $b_{gal} \ge  45^{\circ}$;
Patch B - $193^{\circ} \le \ell_{gal} \le 228^{\circ}$, $b_{gal} \le -45^{\circ}$;
Patch C - $ 65^{\circ} \le \ell_{gal} \le 180^{\circ}$, $b_{gal} \ge  45^{\circ}$.
Being interested in the diffuse Galactic
synchrotron emission at 1.4~GHz, we
subtracted discrete sources (DSs) from the total
intensity map (very few sources are visible
in polarization out of the plane).
We
performed a 2-dimensional Gaussian fitting using the NOD2-software library (\cite{haslam74}),
which also allowed us to estimate the diffuse background.
We are confident that at $|b_{gal}| \gsim 45^{\circ}$ DSs with peak flux densities
exceeding $\sim 1$ Jy have been removed that way
(see Burigana et al.~2006 for a map of the subtracted DSs).\\
We then derived the APSs of the
fluctuation fields (i.e., after subtraction of the mean value
from the maps) for the three selected areas and the whole survey coverage
using the {\tt HEALPix} facility
{\tt anafast}.
Figure~\ref{lowTpatchaps1} shows the APSs obtained
for the selected areas  in both temperature and polarization.

\begin{figure}
\hskip +0.7cm
  \centering
  \includegraphics[width=3.7cm,height=8.1cm,angle=90]{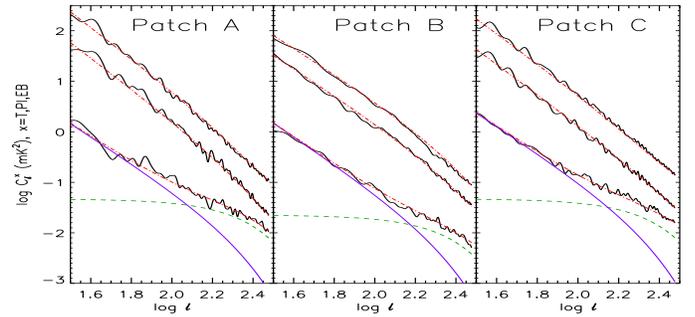}
\vskip +0.2cm
  \caption{APSs
  of the selected cold regions and corresponding best-fit curves (dot-dashed lines).
  From the bottom: $C_{\ell}^{TI}$, $20 \cdot C_{\ell}^{PI}$ and
  $50 \cdot C_{\ell}^{E;B}$.
  In the case of $C_{\ell}^{TI}$, $C_{\ell}^{synch} e^{-(\sigma_{b}\ell)^2}$
  and $c_{\ell}^{src} e^{-(\sigma_{b}\ell)^2}$ are also displayed.}
  \label{lowTpatchaps2}
\end{figure}

\begin{table}
\begin{center}
\begin{tabular}{|c|c|c|c|c|}
\hline
\multicolumn{2}{|c|}{Coverage} & \multicolumn{3}{|c|}{Best-fit parameters}\\
\hline
\multicolumn{2}{|c|}{     } &  ${\rm log}_{10}k$ (mK$^2$) & $\alpha$ & $c^{src}$ (mK$^2$) \\
\hline
A & $C_{\ell}^{TI}$  &  $4.067^{-1.027}_{+0.682}$
        & $-2.60^{+0.60}_{-0.40}$
        & $0.046^{-0.027}_{+0.016}$ \\
        & $C_{\ell}^{PI}$ & $5.000^{-0.420}_{+0.200}$
	& $-3.02^{+0.22}_{-0.11}$ 	
	& $0.002^{-0.002}_{+0.002}$\\
        & $C_{\ell}^{E;B}$ & $5.2550^{-0.305}_{+0.324}$
	& $-3.05^{+0.15}_{-0.17}$
	& $0.003^{-0.003}_{+0.001}$\\
\hline
B & $C_{\ell}^{TI}$  & $4.239^{-0.611}_{+0.016}$
        & $-2.71^{+0.34}_{-0.01}$
        & $0.023^{-0.013}_{+0.008}$\\
        & $C_{\ell}^{PI}$ & $4.146^{-0.105}_{+0.553}$
	& $-2.62^{+0.07}_{-0.35}$
	& $0.003^{-0.003}_{+0.005}$\\
        & $C_{\ell}^{E;B}$ & $4.041^{-0.112}_{+0.405}$
	& $-2.55^{+0.06}_{-0.30}$
	& $0.003^{-0.003}_{+0.005}$\\
\hline
C & $C_{\ell}^{TI}$  & $4.339^{-0.297}_{+0.284}$
        & $-2.64^{+0.03}_{-0.01}$
        & $0.047^{-0.030}_{+0.013}$\\
        & $C_{\ell}^{PI}$ & $4.748^{-0.350}_{+0.650}$
	& $-2.94^{+0.20}_{-0.36}$
	& $0.004^{-0.004}_{+0.004}$\\
        & $C_{\ell}^{E;B}$ & $4.954^{-0.301}_{+0.444}$
	& $-2.94^{+0.20}_{-0.26}$
	& $0.005^{-0.005}_{+0.009}$\\
\hline
\end{tabular}
\end{center}
\vskip -0.2cm
\caption{Least-square best-fit parameters
describing
the APSs of the selected patches in the range $30 \le \ell \le 300$.
	 The errors are given so that the upper (resp. lower) values correspond to the
	 best-fit parameters set with the flattest (resp. steepest) slope.
         }
\label{BFtab}
\end{table}

In all cases,
the polarization APSs
are rather similar\footnote{
In particular, $C_{\ell}^{E} \simeq C_{\ell}^{B}$.
We then introduce the quantity
$C_{\ell}^{E;B}=0.5\cdot(C_{\ell}^{E}+C_{\ell}^{B})$, which will be
used in the following best-fit analysis.
On the contrary, the $E$ and $B$ modes of the primordial CMB anisotropies
differ largely as they are induced by different mechanisms
 (e.g., Seljak \& Zaldarriaga~1997; Kosowsky~1999).}.
This fact might indicate that depolarization due
to differential Faraday rotation should not be relevant in such sky
regions
at the investigated angular scales.
In fact, rotation measure (RM) maps (\cite{johnston}; \cite{dineen})
obtained interpolating RM data of extragalactic
sources show very low values in correspondence to such areas.
However, the degree of polarization
 is on average a few percent,
well below the maximum theoretical value of $\sim 75\%$ (\cite{ginzburg65}).
The reason for the low fractional polarization is not clear. One possibility
is that depolarization other than
differential depolarization is present (e.g.,~\cite{sokoloff}).

The maps show the Galactic synchrotron
emission
and non-subtracted DSs convolved
with the telescope beam and contaminated by noise.
We then fitted the APSs of both total intensity and polarization
(namely $C_{\ell}^{TI}$, $C_{\ell}^{PI}$, and
$C_{\ell}^{E;B}$)
by modelling them as
$C_{\ell} = (C_{\ell}^{synch}+ c^{src})\cdot W_{\ell} + c^{WN}$.
We exploited the power law approximation
$C_{\ell}^{synch} = \kappa \cdot \ell^{~\alpha} \, $
and assumed a symmetric Gaussian beam, i.e., a window function
$W_{\ell}={\rm e}^{-(\sigma_{b} \ell)^2}$, where
$\sigma_b= { \theta_{HPBW} ({\rm rad}) / {\sqrt{8 {\rm ln} 2}} }$~.
The contribution of non-subtracted DSs has been simply
modelled with a constant
term according to the formalism of
Poisson fluctuations from extragalactic point sources (\cite{toffolatti98}),
expected to represent
the largest fraction of the non-subtracted DSs.
From extragalactic source counts at 1.4~GHz
(\cite{prandoni01}) we estimate $c^{src} \simeq 0.09$~mK$^2$
for $C_\ell^{TI}$
(adding or subtracting quoted $1\sigma$ errors to the best-fit number counts
we (conservatively)
get $c^{src} \simeq [0.05 - 0.3]$~mK$^2$).
A guess of the noise contribution is given by
$C_{\ell}^{WN} \sim 4\pi \cdot \sigma_{pix}/N_{pix}$,
where $N_{pix}$ is the number of pixels in the {\tt HEALPix} map and
$\sigma_{pix}=\sigma_{pix,{\tt HEALPix}} \sim N^{-1/2} \cdot \sigma_{pix,{\tt ECP}}$,
$N$ being the number of the {\tt ECP} pixels corresponding to each
{\tt HEALPix} pixel at a given
latitude
(for the total intensity maps of the selected areas,
$\sigma_{pix,{\tt HEALPix}} \sim 13$ mK corresponding
to $c^{WN}\sim 0.003$ mK$^{2}$).
The best-fit results
for
all patches
are listed in Table~\ref{BFtab}
and shown in
Fig.~\ref{lowTpatchaps2}.
For temperature, the parameter range for the Galactic synchrotron emission APS
is $\alpha \sim [-2.71,-2.60]$ for the slope
and ${\rm log}_{10}\kappa \sim [4.067,4.339]$ ($\kappa$ in mK$^{2}$)
for the amplitude.
The DS contribution is in the range $c_{src} \sim [0.023,0.047]$~mK$^{2}$,
which is consistent with the above source counts
estimate, within the errors\footnote{The bulk of the factor
$\sim 2$ of discrepancy between the value of
$c^{src}$ recovered by our fit and that predicted
from best-fit number counts can in fact be produced by the survey
sky sampling ($\theta_{s}$) of $15'$.
For example, considering a lower limit corresponding
to $\simeq 15'/\sqrt{2}$ (instead of 0)
in the integral over $\psi$ in Eq.~(2) of
Toffolatti et al. (1998),
$c_{src}$ decreases by a factor $\simeq 1.53$.}.
For polarization, the derived slopes are generally steeper,
varying in the interval $\alpha \sim [-3.02,-2.62]$
for $C_{\ell}^{PI}$ and $\alpha \sim [-3.05,-2.55]$
for $C_{\ell}^{E;B}$.
The DS contribution is on average much lower
than in temperature,
since it is compatible with zero\footnote{The best-fit results may suggest
a polarization degree (obtained considering
the contribution of the subtracted DSs, $\sim 0.05 - 0.2$~mK$^2$,
to the temperature APS also) considerably higher
than
$\sim 2$\%,
the value found for NVSS extragalactic (mainly steep spectrum)
sources
(Mesa et al.~2002; Tucci et al.~2004).
It may imply
 a presence (or a combination) of spurious instrumental polarization
at small scales,
of a significant contribution from highly polarized Galactic sources
(\cite{manchester98})
non-subtracted in the maps, or of a flattening
of the diffuse synchrotron polarized emission APS at
$\ell \gsim 200-250$
in higher resolution data on smaller sky areas
(Baccigalupi et al.~2001; Carretti et al.~2006).}.

\section{Discussion: implications for CMB observations}

\begin{figure}
\hskip -0.5cm
   \includegraphics[width=9.8cm,height=6.cm]{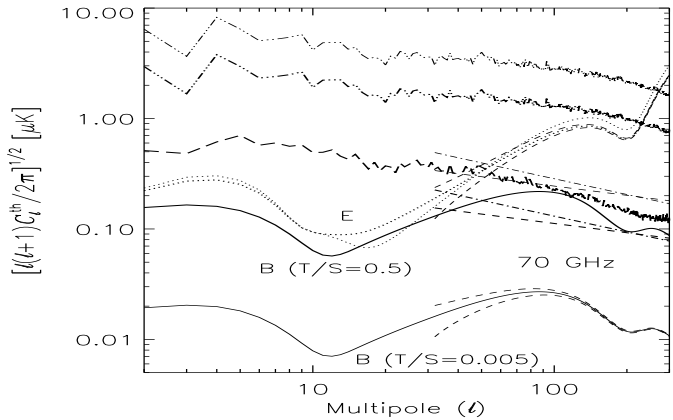}
   \caption{Comparison between the $E$ and $B$ modes of the CMB anisotropy
    at 70 GHz and of the Galactic polarized synchrotron foreground.
    Dotted and solid lines:
    CMB $E$ and $B$ mode for two different tensor-to-scalar ratios
(thin line $T/S=0.005$, thick line $T/S=0.5$).
    Dashed lines:
    uncertainty associated with the cosmic and sampling variance
    for a sky coverage of $\sim 3\%$
    and a binning of 10\% in $\ell$.
    Three dots-dashes:
    average of the $E$ and $B$ modes extrapolated
    from the whole DRAO survey for
   spectral indices $\beta = -2.8$ and $-3$, respectively.
    The {\tt anafast} results have been divided by the window
    function
    to correct for beam smoothing.
    Thick long dashed line: the same as above, but with a
    mask at $|b_{gal}| \le 20$ for $\beta=-3$.
    Dashed (dot-dashed) power law: best-fit result 
    corresponding to Patch~B (resp.~C), rescaled in frequency as above.
    The results are here in terms of thermodynamic temperature.
}
  \label{fig:eb}
 \end{figure}

We have extrapolated the recovered
APS of the $E$ and $B$ modes
to 70~GHz and compared it with the APS of the CMB polarization
anisotropy\footnote{We
used the {\tt CMBFAST} code (version 4.5.1)
for the computation of the
CMB APS (http://www.cmbfast.org/).} for a $\Lambda$CDM model including scalar and tensor
perturbations compatible with the recent WMAP 3-yr
results\footnote{http://lambda.gsfc.nasa.gov/
}
(Spergel et al. 2006).
The frequency range between 60~GHz and 80~GHz seems
the less contaminated by Galactic synchrotron and dust foregrounds in both
temperature (Bennett et al. 2003) and polarization (Page et al. 2006)
at angular scales $\gsim 1^\circ$.
The accurate measure of the $E$ mode is of particular relevance
for breaking the existing degeneracy in cosmological parameter estimation,
when only temperature anisotropy data are available (e.g.,
Bond et al.~1995, Efstathiou \& Bond 1999).
The detection of the primordial $B$ mode is of fundamental importance
for testing the existence of a stochastic cosmological
background of gravitational waves
(e.g., Knox \& Turner 1994).
The results of our comparison have been displayed in Fig.~\ref{fig:eb}
for two choices of the temperature spectral index
($\beta = -2.8, -3$).
The APS extrapolated from the entire DRAO survey is also shown. It
exceeds the CMB $E$ mode even at the
peak at $\ell \sim 100$  for $\beta = -2.8$. For
$\beta = -3$, the two APSs are almost comparable.
Figure~\ref{fig:eb} also shows the APS extrapolated
from the DRAO survey excluding the region $|b_{gal}| \le 20^{\circ}$
and adopting $\beta = -3$\footnote{This APS is consistent with
the WMAP foreground polarization APS at 63~GHz
(see Fig. 17 in Page et al. 2006),
thus supporting an effective $1.4 - 70$~GHz polarization spectral index
$\beta \simeq -3$ far out of the Galactic plane.}.
Such a sky mask
reduces the Galactic APS below the CMB $E$ mode for
$\ell \gsim 50$.
In this case, the CMB $B$ mode
peak at $\ell \sim 100$ is comparable to (or exceeds) the synchrotron signal
for tensor-to-scalar ratios $T/S \gsim 0.5$.
For the three patches
and $T/S \gsim 0.5$,
the cosmological $B$ mode peak at $\ell \sim 100$ is comparable to (or exceeds)
the Galactic synchrotron signal extrapolated with $\beta \simeq -2.8$, while
it is larger by a factor $\gsim 2$
(in terms of $\sqrt{C_\ell}$)
for $\beta \simeq -3$. Furthermore,
a separation of the Galactic synchrotron polarized signal from the CMB one
with a $\sim 5-10$\% accuracy (in terms of $\sqrt{C_\ell}$) would allow
to detect the CMB $B$ mode peak at $\ell \sim 100$ even for
$T/S \sim 0.005$ if $\beta \simeq -3$.
 Similar results for the
 detection of the B mode peak at $\ell \sim 100$ have been inferred
from 1.4~GHz polarization
 observations of a small region with
 exceptionally low Galactic
foreground
contamination
(\cite{carretti06}), though
 at $\ell \sim {\rm few} \times 100$ the CMB B mode is expected to be
 dominated by the B mode generated by lensing
 (Zaldarriaga \& Seljak 1998).
Finally, we note that for a sky coverage comparable with those
of the considered patches
($\sim 3$\%), the cosmic and sampling variance does not significantly
limit the accuracy of the CMB
mode recovery at $\ell$ larger than some tens.

\begin{figure}
\hskip -0.5cm
   \includegraphics[width=9.8cm,height=6.cm]{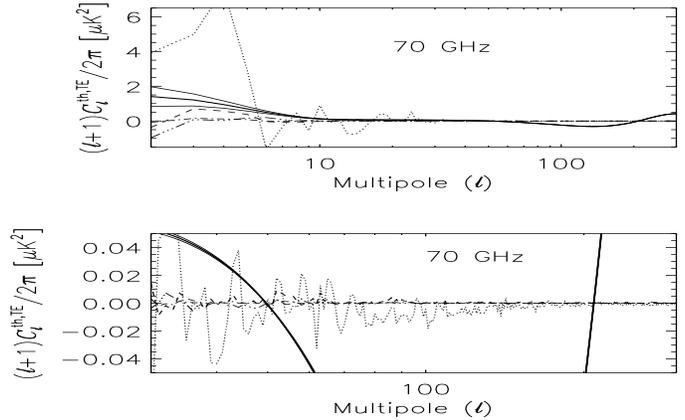}
   \caption{Comparison between the $TE$ modes of the
CMB anisotropy and the Galactic diffuse emission.
Solid lines: CMB $TE$ mode
with $T/S=0.005$ (thick solid line) and corresponding
cosmic variance (region between the thin solid lines),
assuming all sky coverage and without binning in $\ell$.
 Dots:  extrapolated DRAO $TE$ mode
for a spectral index $\beta=-3$.
Dashes (dot-dashes):
as above, but masking the region
at $|b_{gal}| \le 5^\circ$ and adopting
 $\beta = -2.8$ ($\beta =-3$).
Three dots-dashes: as above, but excluding the region
at $|b_{gal}| \le 20^\circ$  ($\beta = -2.8$).
Top and bottom panels are identical, but with a different choice
of the multipole and power range, for a better view of the results.
}
  \label{fig:te}
 \end{figure}

The CMB $TE$ correlation constrains
the reionization history
from the power at low multipoles
and the nature of primordial fluctuations
from the series of peaks and antipeaks
(e.g., Kogut 2003; Page et al.~2006).
In Fig.~\ref{fig:te} we compare the $TE$ mode APS of the Galactic
 emission extrapolated to 70~GHz with the CMB one.
The APS of the whole DRAO survey indicates
a significant Galactic contamination at $\ell \lsim {\rm few} \times 10$,
 even for $\beta \simeq  -3$.
The use of suitable
Galactic
masks (e.g., excluding
regions at $|b_{gal}| \le 5^\circ$ for $\beta\sim-3$
or $|b_{gal}| \le 20^\circ$ for $\beta\sim-2.8$)
greatly reduces the Galactic
foreground
contamination even at low multipoles.
For the three cold patches, the Galactic $TE$ mode
is
negligible compared to the CMB one, independently
of
the adopted
$\beta$.
The $TE$ mode antipeak at $\ell \sim 150$ turns out to be
very weakly affected by Galactic synchrotron contamination in all
cases.

\begin{acknowledgements}
We warmly thank G.~De~Zotti and L.~Toffolatti for useful discussions.
We are grateful to R.~Wielebinski for a careful reading of the original
manuscript.
We warmly thank the anonymous referee for
constructive comments.
Some of the results in this paper have been derived using HEALPix
(G\'orski et al. 2005). The use of the {\tt CMBFAST} code
(version 4.5.1) is acknowledged. L.L.P. was supported for this research
through a stipend from
the International Max Planck Research School (IMPRS)
for Radio and Infrared Astronomy at the Universities of Bonn and
Cologne.
\end{acknowledgements}

\end{document}